# High Curie temperature $Mn_5Ge_3$ thin films produced by non-diffusive reaction

E. Assaf, A. Portavoce[*], K. Hoummada, M. Bertoglio, and S. Bertaina

*IM2NP, CNRS/Aix-Marseille University, Faculté des Sciences de Saint-Jérôme case 142, 13397 Marseille, France*

**ABSTRACT**

Polycrystalline $Mn_5Ge_3$ thin films were produced on $SiO_2$ using magnetron sputtering and reactive diffusion (RD) or non-diffusive reaction (NDR). In situ X-ray diffraction and atomic force microscopy were used to determine the layer structures, and magnetic force microscopy, superconducting quantum interference device and ferromagnetic resonance were used to determine their magnetic properties. RD-mediated layers exhibit similar magnetic properties as MBE-grown monocrystalline $Mn_5Ge_3$ thin films, while NDR-mediated layers show magnetic properties similar to monocrystalline C-doped $Mn_5Ge_3C_x$ thin films with $0.1 \leq x \leq 0.2$. NDR appears as a CMOS-compatible efficient method to produce good magnetic quality high-curie temperature $Mn_5Ge_3$ thin films.

The ferromagnetic compound $Mn_5Ge_3$ is expected to allow spin current injection in Ge [1-4]. Consequently, the use of $Mn_5Ge_3$ thin films as ferromagnetic contacts on the source and drain of Ge-based transistors is currently investigated in order to produce spintronic devices [4-6]. In the literature, $Mn_5Ge_3$ films are generally obtained via reactive diffusion, reproducing the "Self-aligned silicide process" used in the Si Complementary metal oxide semi-conductor (CMOS) technology to produce ohmic contacts on transistor's source, drain and gate [5-7]. Usually, an Mn film is deposited on an Ge substrate by molecular beam epitaxy (MBE) [6,8] before to be in-situ thermally annealed under ultra-high vacuum, in order to form a $Mn_5Ge_3$ layer at the Mn/Ge interface. The control of the annealing conditions (time, temperature) allows the $Mn_5Ge_3$ layer to grow until the complete consumption of the initial Mn layer, without the formation of the next compound in the phase formation sequence $Mn_{11}Ge_8$. The obtained $Mn_5Ge_3$ layer can be monocrystalline in epitaxy on Ge if the Mn film is deposited on a Ge(111) substrate [6,8]. This self-aligned germanide process has the benefit to be compatible with industrial CMOS processes [5]. However, the Curie temperature ($T_c$) of $Mn_5Ge_3$ (~ 297 K) needs to be increased in order to obtain spin current injection at temperatures higher than room temperature (RT) [5,9-10]. Usually, this is achieved using C [7,10-14] or Fe doping [15], the dopant being co-evaporated with Mn on the Ge substrate, complicating the deposition process. Furthermore, ultra-high vacuum MBE growth is not currently used in CMOS process lines, due to its significant production cost. Metallic films used for ohmic contact fabrication in CMOS technology are deposited by magnetron sputtering. In addition, one of the central drawbacks of $Mn_5Ge_3$ films grown on Ge is the weak stability of the film, which can transform into $Mn_{11}Ge_8$ during a following annealing in following transistor fabrication process steps [5]. Though, electrical spin injection in *n*-type Ge through a 2 nm-thick $Al_2O_3$ tunneling barrier was recently demonstrated, from a 40-nm thick $Mn_5Ge_3C_{0.8}$ polycrystalline layer grown by Mn, Ge, and C simultaneous magnetron sputtering at 400 °C [4].

In this work, the magnetic properties of $Mn_5Ge_3$ films produced by means of two different methods compatible with CMOS technology were investigated by magnetic force microscopy (MFM), superconducting quantum interference device (SQUID) and Ferromagnetic resonance (FMR). The magnetron sputtering technique [4,9] is shown to produce $Mn_5Ge_3$ films exhibiting similar magnetic properties as $Mn_5Ge_3$ films produced by MBE. Furthermore, non-diffusive reaction (NDR) from an amorphous $Mn_{0.625}Ge_{0.375}$ film leads to the formation of $Mn_5Ge_3$ films exhibiting higher $T_c$ and better magnetic properties than $Mn_5Ge_3$ films produced by reactive diffusion (RD) of Mn and Ge layers deposited in same conditions.

Two types of samples were prepared using a commercial magnetron sputtering system exhibiting a base pressure of $10^{-8}$ Torr and allowing simultaneous sputtering of three different targets. All the layers were deposited at RT in same conditions (Ar gas pressure…) using a 99.9999% pure Ar gas flow to sputter a 99.99% pure Ge target and a 99.9% pure Mn target. The sputtered layers were deposited on a $SiO_2$ layer grown on an Si(001) substrate. The $SiO_2$ layer acts as a diffusion barrier, preventing any intermixing between the layers and the Si substrate. For RD experiments, a $31 \pm 2$ nm-thick polycrystalline Mn layer was deposited on top of a $202 \pm 13$ nm-thick amorphous Ge layer. A sputtered Ge substrate was preferred to a commercial Ge wafer, allowing the results obtained with this sample to be fully comparable with the results obtained with the second type of sample (sputtered Ge in the two cases). For NDR experiments, a $45 \pm 4.5$ nm-thick amorphous $Mn_{0.625}Ge_{0.375}$ film was directly deposited on the $SiO_2$ layer. These two types of samples were in situ annealed under vacuum ($10^{-6}$ Torr) in an X-ray diffraction (XRD) setup following a heating ramp made of 5 °C per minute steps separated by 4 minutes-long XRD measurements at constant temperature ($T$), corresponding to an average heating ramp of 1 °C min$^{-1}$. The XRD measurements were performed between RT and 400 °C in the Bragg-Brentano geometry, using a Cu K$_\alpha$ source ($\lambda_{K\alpha}$ = 0.154 nm). The thicknesses of the films were measured before and after annealing using the X-ray reflectivity technique. The thickness ($t$) of the Mn$_5$Ge$_3$ film in contact with Ge obtained by RD was found to be $t$ ~ 48 ± 5 nm, while the thickness of the Mn$_5$Ge$_3$ film in contact with SiO$_2$ obtained by NDR was found to be $t$ ~ 47 ± 2.5 nm. The sample surface topography was studied using a NT-MDT SMENA atomic force microscope (AFM) in air in non-contact mode. The magnetization component perpendicular to the sample surfaces was studied at RT by MFM in air, while the magnetic properties of the films were studied using SQUID and FMR measurements performed at RT and at 200 K with a conventional Bruker EMX spectrometer operating at $f$ = 9.62 GHz.

In situ XRD measurements (not presented here) showed the same phase formation sequence as in the case of RD between a polycrystalline Mn film and an amorphous Ge layer deposited by e-beam evaporation [5]. Only two phases are observed in the sequence, the first phase being Mn$_5$Ge$_3$ and the second phase being Mn$_{11}$Ge$_8$. However, in the case of e-beam evaporation, Mn$_5$Ge$_3$ was found to form at $T$ ~ 210 °C and to transform into Mn$_{11}$Ge$_8$ at $T$ ~ 310 °C [5], while in the present case, Mn$_5$Ge$_3$ was found to form at ~ 240 °C and to be stable up to temperatures as high as 400 °C. In the case of Mn$_5$Ge$_3$ formation by NDR, in situ XRD measurements showed only the formation of Mn$_5$Ge$_3$ at $T$ ~ 225 °C. Fig. 1 presents the diffractograms obtained at RT after annealing the RD (a, red line) and NDR (b, black line) samples up to 400 °C. For the RD sample, three diffraction peaks corresponding to Ge ($2\theta$ = 27.5°, 45.6°, and 53.9° corresponding respectively to the Ge planes (111), (220), and (311)) are easily detected (triangles in fig. 1). They correspond to the 200 nm-thick Ge layer that crystallized during annealing [5]. In addition, six diffraction peaks at $2\theta$ = 30.9°, 35.8°, 38.7°, 42.7°, 43.8°, and 56.7°, corresponding respectively to the atomic planes (111), (002), (210), (211), (112), and (311) of $Mn_5Ge_3$ are also observed. No peak belonging to $Mn_{11}Ge_8$ can be detected [5]. For the NDR sample, the same six diffraction peaks corresponding to $Mn_5Ge_3$ are also observed, with an additional peak at $2\theta = 46.3°$, corresponding to the $Mn_5Ge_3$(202) planes. The relative intensity of the $Mn_5Ge_3$ peaks is different in the two samples. For example, the crystallographic direction exhibiting the maximum intensity in the RD sample is the (211), while it is the (002) direction in the NDR sample. Thus, the two $Mn_5Ge_3$ layers do not possess the same texture. Fig. 2 presents AFM images of the sample surfaces. The grains are easily observed due to grain boundary (GB) grooving that occurred during annealing, resulting from equilibrium between the surface and interface energies at GBs [16-19]. For example, fig. 3a presents a one-dimensional profile (1D) measured on the NDR sample surface by AFM. GB grooves can exhibit a large width in the two samples (> 150 nm). However, their depth is about 3 nm that is negligible compared to the $Mn_5Ge_3$ layer thickness (~ 50 nm). The average grain width ($L$) was determined considering that the lateral size of a given grain corresponds to the distance between the bottoms of the grooves located on each side of the grain [16-19], as presented in fig. 3a. $L$ was averaged over all the grains detected on AFM images of $25 \times 25$ μm$^2$ and $15 \times 15$ μm$^2$ for the RD and NDR samples, respectively [20-21]. $L \sim 1.09$ μm and the surface roughness (root mean squared) $r \sim 1.5$ nm for the $Mn_5Ge_3$ layer produced by RD, while $L \sim 0.4$ μm and $r \sim 1.2$ nm for the $Mn_5Ge_3$ layer produced by NDR. The Scherrer equation [22] applied to the XRD data of the RD- and NDR-mediated layers gives for the grain size in the direction perpendicular to the surface $d = 51.6$ nm and $d = 47.7$ nm, respectively. Thus, the two layers present a similar surface roughness, and in both cases, the grains are columnar ($t \sim d$) and their lateral size is larger than the $Mn_5Ge_3$ layer thickness ($L \geq t$). Though, $Mn_5Ge_3$ grains are about two times larger in the layer obtained by RD. Fig. 4a and 4b present the MFM images (phase−$\varphi$ contrast) obtained on the sample surfaces at RT without external magnetic field. Magnetic domains are visible on the two samples, showing that the samples are ferromagnetic at RT. However, the magnetic domains are more easily detected on the NDR sample (fig. 3b), with a maximum contrast $\Delta\varphi \sim 1.7°$, than on the RD sample, with a maximum contrast $\Delta\varphi \sim 1.4°$. The domains do not follow the sample surface topologies. The average domain width is found to be $\sim 0.65 \pm 0.2$ μm and $\sim 0.39 \pm 0.1$ μm, and the average domain length is found to be $\sim 1.5 \pm 0.2$ μm and $\sim 0.9 \pm 0.1$ μm, on the RD and NDR samples, respectively. The angular ($\alpha$) variation of the resonance field $Hr$ is shown in fig. 4c. These measurements confirm that due to the film geometry, the samples possess a shape anisotropy with an easy axis in the direction parallel to the surface ($\alpha = 0°$) and a hard axis in the direction perpendicular to the surface ($\alpha = 90°$). Despite that the samples were produced using the same deposition technique in same conditions, the $Mn_5Ge_3$ layer produced by NDR exhibits a resonance field difference $\Delta Hr$ between the hard ($Hr_\perp$) and easy axis ($Hr_{//}$) significantly higher than in the case of the $Mn_5Ge_3$ layer produced by RD. In addition, the results obtained on the RD sample are similar to the results obtained on a monocrystalline $Mn_5Ge_3$ layer in epitaxy on Ge(111) produced by RD of an Mn layer deposited by MBE on a Ge(111) substrate [14]. For the latter, $Hr_{//}$ and $Hr_\perp$ were found to be about $\sim 2.7$ and $5.0$ kOe, respectively, giving $\Delta Hr \sim 2.3$ kOe at RT [14]. In our case, $Hr_{//} \sim 3.1$, $Hr_\perp \sim 4.8$, and $\Delta Hr \sim 1.7$ kOe for the polycrystalline $Mn_5Ge_3$ layer obtained by RD. $\Delta Hr$ depends on the ferromagnetic layer $T_c$. For example, $\Delta Hr \sim 2.3$ kOe was shown to correspond to $T_c = 315 \pm 10$ K for the epitaxial MBE-mediated $Mn_5Ge_3$ layer [14]. This is in agreement with the expected $T_c \sim 297$ K of the $Mn_5Ge_3$ compound, as well as with our MFM measurements (fig. 4a) showing a faint signature of ferromagnetic domains at RT. Contrasting with $Mn_5Ge_3$ epitaxied layers [14], the FMR measurements can be well simulated using the Chappert model (solid lines in fig. 4c) containing only two magnetic anisotropies [23]: the shape and magnetocrystalline anisotropies. This is probably due to the polycrystalline nature of the present $Mn_5Ge_3$ films. Using the RT $Mn_5Ge_3$ saturation magnetisation $M = 360$ emu cm$^{-3}$ [24], the magnetocrystalline anisotropy constants ($K_{1\perp}$ and $K_{2//}$) of the polycrystalline RD-mediated Mn$_5$Ge$_3$ layer were calculated considering a gyromagnetic ratio $\gamma/2\pi = 2.8$ GHz kOe$^{-1}$ [23]. We found $K_{1\perp} = 0.57 \times 10^6$ erg cm$^{-3}$, and $K_{2//}$ = $5.42 \times 10^4$ erg cm$^{-3}$. FMR measurements compared to SQUID measurements performed on C-doped epitaxial MBE-mediated Mn$_5$Ge$_3$C$_x$ layers showed that $\Delta Hr \sim 3.7$ kOe ($Hr_{//} \sim 2.3$ and $Hr_\perp \sim 6.0$ kOe) for $x = 0.1$ corresponds to $T_c = 345 \pm 10$ K, and $\Delta Hr \sim 7.0$ kOe ($Hr_{//} \sim 1.6$ and $Hr_\perp \sim 8.6$ kOe) for $x = 0.2$ corresponds to $T_c = 450 \pm 30$ K [14]. $Hr_{//} \sim 2.0$, $Hr_\perp \sim 7.5$, and $\Delta Hr \sim 5.5$ kOe in the polycrystalline Mn$_5$Ge$_3$ layer obtained by NDR (fig. 4c), which suggests that the $T_c$ in the NDR-mediated Mn$_5$Ge$_3$ layer is larger than in the RD-mediated layer, and is comprised between 345 and 450 K. SQUID measurements were performed on the samples in order to confirm the FMR measurements. As expected, the magnetization variations versus temperature presented in fig. 5 show that the Curie temperature is $\sim 300$ K and $\sim 385$ K, in the RD- and NDR-mediated Mn$_5$Ge$_3$ films, respectively. FMR measurements were also performed at 200 K (not shown here). However, the signal from the RD-mediated layer being too broaden (red open symbols in the inset in fig. 4c), measurements could be performed only on the NDR-mediated Mn$_5$Ge$_3$ layer (black solid symbols in the inset in fig. 4c). We measured $Hr_{//} \sim 1.3$ kOe, $Hr_\perp \sim 9.1$ kOe, and $\Delta Hr \sim 7.8$ kOe in this layer at 200 K. The significant FMR signal difference at 200 K between the two Mn$_5$Ge$_3$ layers presented in the inset in fig. 4c shows clearly that the NDR-mediated layer exhibits better magnetic properties than the RD-mediated layer [25].

$T_c$ in Mn$_5$Ge$_3$ is mainly related to Mn-Mn atoms interactions [26], and thus, depends on the distance between Mn atoms in the compound. This distance can be modified by stress and/or alloying with a third element. The Mn$_5$Ge$_3$ bulk compound possess a hexagonal structure, with the standard (calculated) lattice parameters a = b = 0.7185 nm and c = 0.5053 nm [27]. These parameters were measured in the samples from the XRD data presented in fig. 1. For the RD mediated layer, the lattice parameters were found to be a = b = 0.7158 nm and c = 0.5014 nm. For the NDR mediated layer, the lattice parameters were found to be a = b = 0.7135 nm and c = 0.5016 nm. These data are very similar, the difference between the measurements being in the sub-angstrom range ($10^{-2}$ angstrom), it is related to measurement accuracy. Consequently, the $T_c$ variations observed between the RD- and NDR-mediated samples cannot be explained by stress effect. In our case, there is no alloying effect, since Ge and Mn were the only elements sputtered on the substrates. However, sputtered layers are known to contain C and O atoms with concentrations that can reach several percent depending on the purity of the Ar gas and of the targets, as well as on the deposition rate and of the setup base pressure. The C doping effect allowing for $T_c$ increase in $Mn_5Ge_3$(C) was shown to be of electronic nature, interstitial C atoms in the $Mn_5Ge_3$ compound promoting the appearance of the 90° ferromagnetic superexchange between Mn atoms [26]. Thus, the reason for the $T_c$ difference between the two $Mn_5Ge_3$ layers could be related to a difference of C incorporation, related to the two reaction types RD and NDR. Usually, RD allows impurities such as C and O to be pushed-out to interfaces and to GBs during the layer growth, allowing the formation of thin films significantly less contaminated than the initial sputtered layers [28]. At the opposite, atomic diffusion being not necessary during NDR [29], C incorporation could be more effective during this process, leading to a $T_c$ increase in NDR-mediated $Mn_5Ge_3$ layers.

In conclusion, $Mn_5Ge_3$ layers were grown on $SiO_2$ using thin film magnetron sputtering followed by thermal annealing up to 400 °C. Two different reaction types were investigated: RD and NDR. Magnetron sputtering is shown to allow polycrystalline $Mn_5Ge_3$ layers, exhibiting similar magnetic properties as MBE-mediated monocrystalline layers in epitaxy on Ge(111), to be produced. RD-mediated $Mn_5Ge_3$ layers exhibit magnetic properties matching usual $Mn_5Ge_3$ properties, while NDR-mediated $Mn_5Ge_3$ layers show similar magnetic properties as C-doped $Mn_5Ge_3C_x$ layers with $0.1 \leq x \leq 0.2$ and $T_c \sim 385$ K. The difference of magnetic properties between the layers produced by RD and NDR could be related to a difference of C incorporation in the $Mn_5Ge_3$ layers, RD leading to a layer purification process decreasing C incorporation and, at the opposite, NDR supporting C incorporation. The NDR process appears as an interesting technique to produce low-cost high-Curie temperature $Mn_5Ge_3$ layers.

This work was supported by the French government through the program "Investissements d'Avenir A*MIDEX" (Project APODISE, no. ANR-11-IDEX-0001-02) managed by the National Agency for Research (ANR).

# REFERENCES

[1] S. Picozzi, A. Continenza, and A.J. Freeman, Phys. Rev. B **70**, 235205 (2004).

[2] R.P. Panguluri, C. Zeng, H.H. Weitering, J.M. Sullivan, S.C. Erwin, and B. Nadgorny, Phys. Stat. Sol. (b) **242**, R67 (2005).

[3] Yu.S. Dedkov, M. Holder, G. Mayer, M. Fonin, and A.B. Preobrajenski, J. Appl. Phys. **105**, 073909 (2009).

[4] I.A. Fischer, L.-T. Chang, C. Sürgers, E. Rolseth, S. Reiter, S. Stefanov, S. Chiussi, J. Tang, K.L. Wang, and J. Schulze, Appl. Phys. Lett. **105**, 222408 (2014).

[5] O. Abbes, A. Portavoce, V. Le Thanh, C. Girardeaux, and L. Michez, Appl. Phys. Lett. **103**, 172405 (2013).

[6] M. Petit, L. Michez, C.-E. Dutoit, S. Bertaina, V.O. Dolocan, V. Heresanu, M. Stoffel, V. Le Thanh, Thin Solid Films **589**, 427 (2015).

[7] L.-A. Michez, F. Virot, M. Petit, R. Hayn, L. Notin, O. Fruchart, V. Heresanu, M. Jamet, and V. Le Thanh, J. Appl. Phys. **118**, 043906 (2015).

[8] C. Zeng, S.C. Erwin, L.C. Feldman, A.P. Li, R. Jin, Y. Song, J.R. Thompson, and H.H. Weitering, Appl. Phys. Lett. **83**, 5002 (2003).

[9] E. Sawatzky, J. Appl. Phys. **42**, 1706 (1971).

[10] A. Spiesser, V. Le Thanh, S. Bertaina, and L.A. Michez, Appl. Phys. Lett., Vol. **99**, 121904 (2011).

[11] M. Gajdzik, C. Sürgers, M.T. Kelemen, and H.v. Löhneysen, J. Magnetism and Magnetic Materials **221**, 248 (2000).

[12] I. Slipukhina, E. Arras, Ph. Mavropoulos, and P. Pochet, Appl. Phys. Lett. **94**, 192505 (2009).

[13] A. Spiesser, I. Slipukhina, M.-T. Dau, E. Arras, V. Le Thanh, L. Michez, P. Pochet, H. Saito, S. Yuasa, M. Jamet, and J. Derrien, Phys. Rev. B **84**, 165203 (2011).

[14] C.-E. Dutoit, V.O. Dolocan, M. Kuzmin, L. Michez, M. Petit, V. Le Thanh, B. Pigeau, and S. Bertaina, J. Phys. D: Appl. Phys. **49**, 045001 (2016).

[15] T.Y. Chen, C. L. Chien, and C. Petrovic, Appl. Phys. Lett. **91**, 142505 (2007).

C dopage

[16] E. Rabkin, L. Klinger, T. Izyumova, A. Berner, and V. Semenov, Acta mater. **49**, 1429 (2001).

[17] A. Otsuki, Acta mater. **49**, 1737 (2001).

[18] H. Yoshida, K. Yokoyama, N. Shibata, Y. Ikuhara, T. Sakuma, Acta Materialia **52**, 2349 (2004).

[19] A. Ramasubramaniam, and V.B. Shenoy, Acta Materialia **53**, 2943 (2005).

[20] D. Nečas, P. Klapetek, Cent. Eur. J. Phys. **10**, 181 (2012).

[21] A. Portavoce, K. Hoummada, F. Dahlem, Surface Science **624**, 135 (2014).

[22] C. Eid, E. Assaf, R. Habchi, P. Miele and M. Bechelany, RSC Adv. **5**, 97849 (2015).

[23] C. Chappert, K. Le Dang, P. Beauvillain, H. Hurdequint, and D. Renard, Phys. Rev. B **34**, 3192 (1986).

[24] A. Spiesser, Ph.D. Thesis, Aix-Marseille University, 2011.

[25] W. Platow, A.N. Anisimov, G.L. Dunifer, M. Farle, and K. Baberschke, Phys. Rev. B **58**, 5611 (1998).

[26] I. Slipukhina, E. Arras, Ph. Mavropoulos, and P. Pochet, Appl. Phys. Lett. **94**, 192505 (2009).

[27] L. Castelliz, Monatsh. Chem. **84**, 765 (1953).

[28] K. Hoummada, I. Blum, D. Mangelinck, A. Portavoce, Appl. Phys. Lett. **96**, 261904 (2010).

[29] A. Portavoce, and G. Tréglia, Phys. Rev. B **82**, 205431 (2010).

**FIGURE CAPTIONS**

**FIG. 1**. XRD measurements performed at RT after annealing the samples up to 400 °C using an average heating ramp of 1 °C per minute: a) RD sample, and b) NDR sample.

**FIG. 2.** AFM measurements performed after annealing: a) RD sample, and b) NDR sample.

**FIG. 3.** 1D profiles measured after annealing on the NDR sample surface (on different regions), a) by AFM and, b) by MFM.

**FIG. 4.** Magnetic measurements performed at 300 K, a) $5 \times 5$ µm$^2$ MFM image obtained on the RD sample, b) $5 \times 5$ µm$^2$ MFM image obtained on the NDR sample, and c) angular variation of the resonance field measured by FMR on the RD (open squares) and NDR (solid squares) samples. The red solid lines correspond to simulations using the Chappert model [20]. The inset shows the FMR signal ($\alpha = 90°$) measured at 200 K on the RD (red open squares) and NDR (black solid circles) samples.

**FIG. 5.** Magnetization evolution versus temperature in the RD (open squares) and NRD (solid squares) samples with a magnetic field of 0.01 T applied in-plane. The Curie temperature of the RD-mediated Mn$_5$Ge$_3$ layer is ~ 300 K, while it is ~ 385 K for the NRD-mediated layer (linear approximation). The inset presents the M-H loop at $10 \leq T \leq 270$ K for the NDR sample.

**Figure 1**

E. Assaf et al.

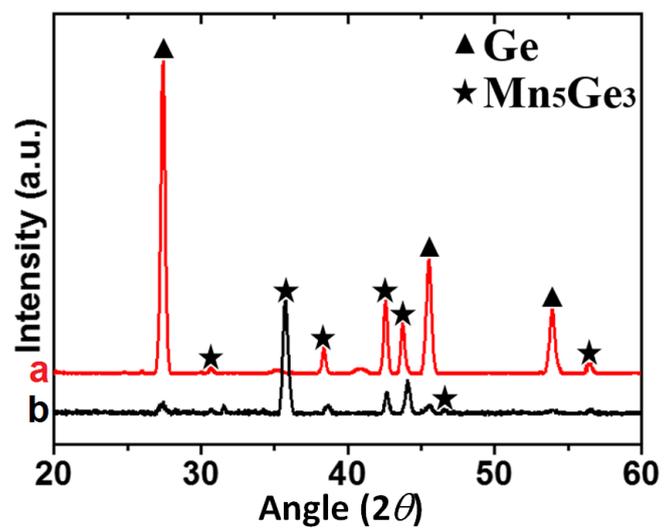

**Figure 2**

E. Assaf et al.

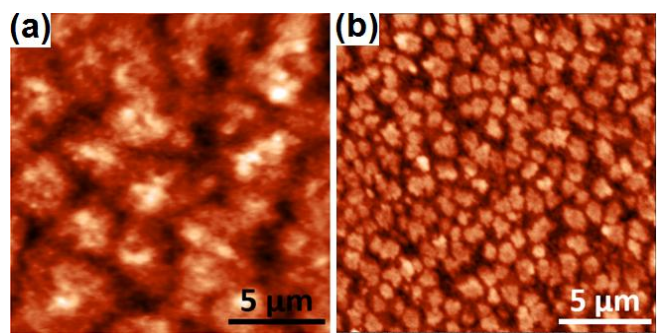

**Figure 3**

**E. Assaf et al.**

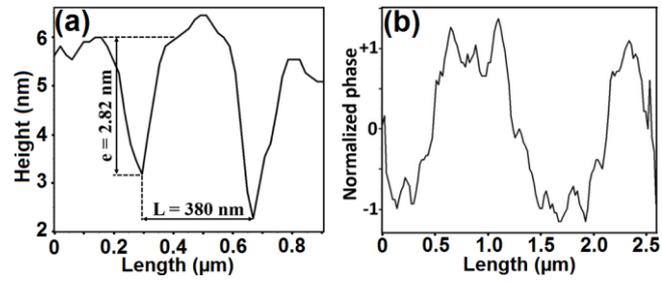

**Figure 4**

**E. Assaf et al.**

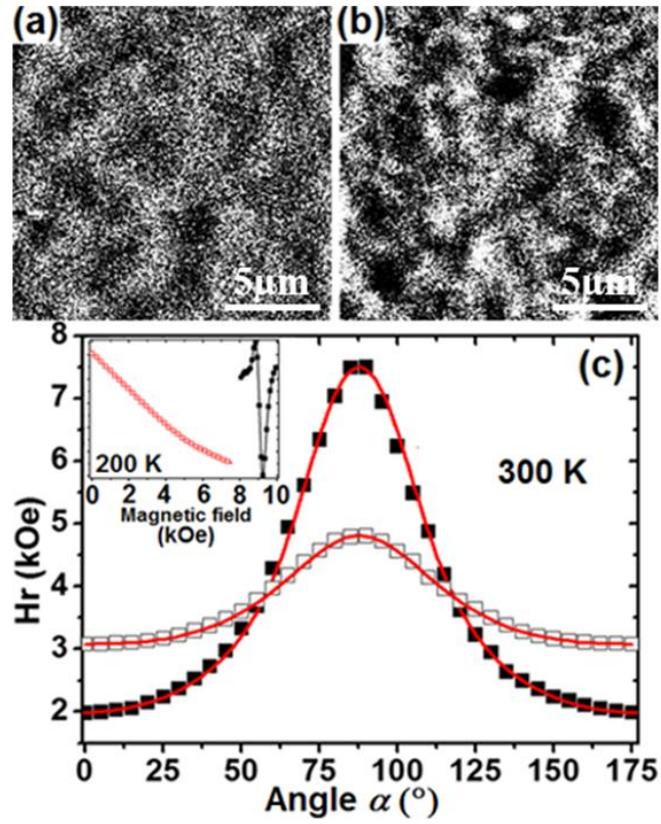

**Figure 5**

E. Assaf et al.

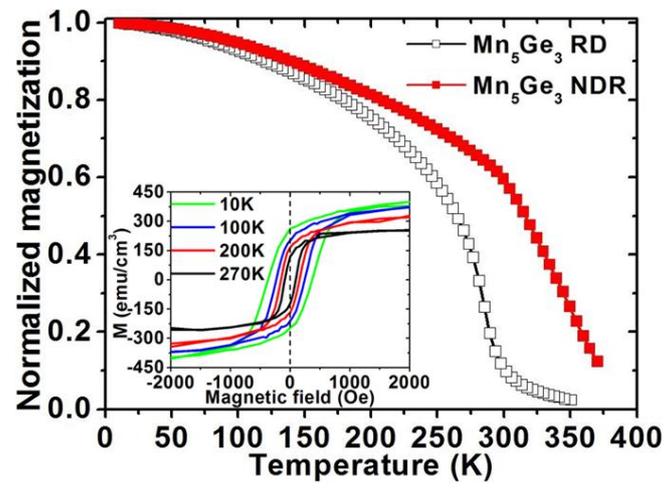